\documentclass[preprint,pre,showpacs,preprintnumbers,amsmath,amssymb,arial]{revtex4}

\usepackage{graphicx}
\usepackage{dcolumn}
\usepackage{bm}
\usepackage{epsfig}
\begin{document}
\title{Influence of the backward propagating waves on the threshold in planar nematic liquid crystal films}
\author{Dmitry O. Krimer}
\affiliation{Max Planck Institute for the Physics of Complex Systems, N\"othnitzer Str. 38, D-01187 Dresden, Germany}
\author{Andrey E. Miroshnichenko}
\affiliation{Nonlinear Physics Centre and Centre for Ultra-high bandwidth Devices for Optical Systems (CUDOS), Australian National University, Canberra ACT 0200, Australia}
\author{Etienne Brasselet}
\affiliation{Centre de Physique Optique Mol\'eculaire et Hertzienne, Universit\'e Bordeaux 1, CNRS, 351 Cours de la Lib\'eration, 33405 Talence Cedex, France}

\date{\today}
\begin{abstract}
We analyze theoretically the influence of backward propagating waves on the primary threshold 
when a linearly polarized light impinges at normal incidence on a planarly aligned nematic liquid crystal films.
We show, that the primary threshold, as a function of the phase delay induced by the nematic layer, exhibits oscillations. The amplitude of oscillations depends strongly on the drop of the refractivity indices of the nematic and outer media at the boundaries.
\end{abstract}

 \maketitle


%
\section{Introduction}
%

Many aspects of complex light-induced nonlinear orientational phenomena in liquid crystals have been investigated during the last few decades, with a particular emphasis on the dynamical richness arising from the strong light-matter coupling associated with the elastic and anisotropic optical properties of ordered mesophases \cite{Tabiryan86_review,krimer_rev,khoo}. From the optical point of view, nematics are uniaxial media with the optical axis along the local average direction of the molecular axis called director, ${\bf n}$. Depending on the boundary conditions and bulk ordering characteristics, a light-driven orientational instability may take place above a threshold intensity usually referred to as the optical Fr\'eedericksz transition (OFT). Besides their fundamental interest, the optically induced phenomena in liquid crystals seem to be finding their way for technological applications. It has been recently proposed \cite{Mirosh_08_1,Mirosh_08_2,MBKK10} to realize all-optical photonic switching devices using dielectric periodic structures infiltrated with homeotropically or planar aligned  nematics. 

When considering a linearly polarized light field impinging at normal incidence onto a planar nematic cell, one should distinguish between two situations when the polarization plane is parallel or perpendicular to the initial director uniform distribution. In both cases, twisted elastic reorientation modes appear above a light intensity threshold. The first one corresponds to the coherent polarization conversion process mediated by two-beam coupling between the incident extraordinary wave and a lower-frequency noisy ordinary component. The latter field component is due to scattering of the incident light by the director orientational fluctuations and experiences gain during the propagation of light throughout the cell  \cite{Khoo,Gusev}. The second situation, which corresponds to an ordinary polarized incident light, is associated with the twist Fr\'eedericksz  transition. Up to now, this particular case has only been considered theoretically because of the predicted threshold for orientational instability is turned out to be very high \cite{Santamato}. Recently, the twist OFT has been revisited and it was shown that the stationary bifurcation solution derived in Ref.~\cite{Santamato} is valid for optically thin nematic layers. Indeed, for thicker films, the twisted OFT takes place via a Hopf bifurcation \cite{krim_planar}, which has not been demonstrated experimentally yet. It is worth noting that the planar geometry has attracted an interest following seemingly experimentally accessible intensity threshold using a periodic structure \cite{Laudyn}.

In this work, in an attempt to guide future experimental investigations we quantitatively estimate the influence of unavoidable back and forward light scattering on the twist OFT threshold that are due to boundary refractive indices mismatch in any real sample. This effect is usually neglected, which is well justified for homeotropic alignment. However, when planar alignment is considered, the perfectly matching condition can be achieved only by utilizing anisotropic substrates. This is obviously an inconvenient solution in practice. Our analysis reveals the crucial role of backward waves in the calculation of the threshold itself.

%
\section{The model}
\label{theor_model}
%

\begin{figure}
\center{\includegraphics[width=0.7\columnwidth ]{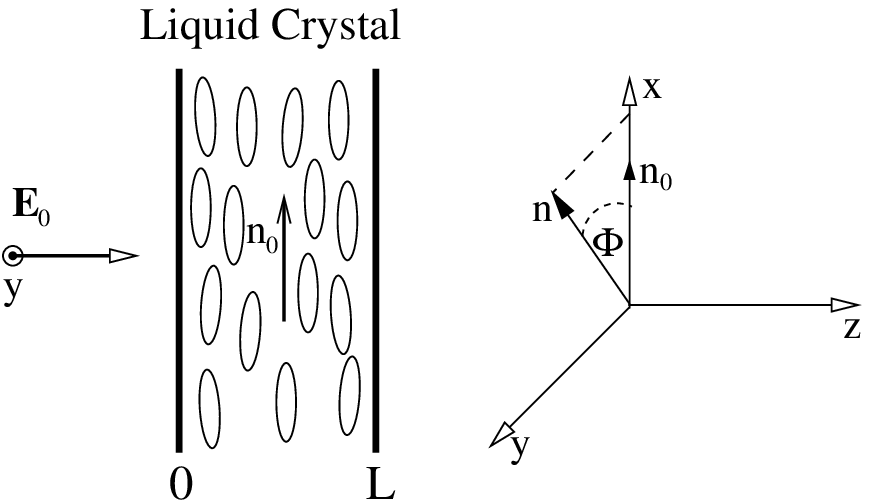}}
\caption{Light-matter interaction geometry: linearly polarized light along the ${\bf y}$ direction incident perpendicularly on a nematic layer with the director ${\bf n_0} \parallel {\bf x} $  (planar state). The components of the director ${\bf n}$ are described in terms of the twist angle
$\Phi$.} \label{fig1_setup}
\end{figure}

We consider a linearly-polarized plane wave with wavelength $\lambda$ impinging at normal incidence (along $z$) on a planarly aligned nematic liquid crystal layer of a thickness $L$ (see  Fig.~\ref{fig1_setup}). The unperturbed director ${\bf n}_0$ lies along the $x$ axis and the light is polarized along the $y$ direction, i.e. we deal with an incident ordinary wave. Theoretical description of optically induced orientational phenomena in nematics is a well-posed problem that consists of coupled nematodynamic and Maxwell's equations for the light propagation inside the anisotropic medium \cite{Gen}. There are several common simplifications that are employed to solve these equations. First of all, fluid flow associated with the director reorientation (backflow effect) is neglected. Indeed, flow plays only a passive role and in most known cases  lead to slight quantitative differences only \cite {krimer_flow,demeter_flow}. In this study we neglect all transverse effects and assume that all variables depend solely on $z$. It allows to simplify drastically the Maxwell's equations. Therefore, the light beam might be treated as a plane wave and any experiment aiming to verify the theoretical predictions should be performed by using a laser beam with a diameter much larger than the thickness of the nematic layer. The applicability of such a one-dimensional approximation has already been validated for homeotropic nematic samples, with ${\bf n}_0 \|{\bf z}$. It turned out that it also works rather well for the beam diameters of the order of $L$ \cite{JOSA_EBDK,JOSA_DKEB}. Importantly, the realization of spatially inhomogeneous transverse structures in the $(x,y)$ plane is hardly probable. As was shown in \cite{Ledney} in the limit of thin layers, it is a case only for  strongly anisotropic ratio between elastic constants which is not a case for common nematics. Next, we assume that the director always keeps planar orientation at the boundaries (strong anchoring conditions), i.e. ${\bf n}(z=0,t) = {\bf n}(z=L,t) = {\bf n}_0$. We introduce the representation adapted to our geometry in terms of the spherical angles $\Theta(z,t)$ and $\Phi(z,t)$  such that ${\bf n}=(\cos \Phi \sin\Theta,\sin \Phi\sin\Theta,\cos\Theta)$. However, as follows from the simulations, the director reorients in the $(x,z)$ plane only for the all the regimes explored in the present study. Therefore $\Theta=\pi/2$ and the director is described by the twist angle $\Phi$ only. Such a situation is intuitively expected because the light should transmit a lot of energy to deform the director across the layer. The strong boundary conditions in terms of the twist angle read
\begin{eqnarray}
\label{bound_cond_Phi} \Phi(z=0,t) = \Phi(z=\pi,t) = 0.
\end{eqnarray}
\section{Primary instability with ideal light field boundary conditions}
\label{subsec_id_bc}

We first neglect the effect of reflection at the boundaries $z=0,\pi$ (ideally matched conditions) and calculate the thresholds by performing the linear stability analysis of the planar state. We eventually obtain the following integro-differential equation for $\Phi$ \cite{krim_planar}
\begin{eqnarray}
\label{dir_eq_F_zero}
\partial_t\Phi\!=\!\partial_z^2\Phi\!+\!2\rho\Delta^2
\left(\Phi\!+\!\Delta\int_0^z\Phi(z')\sin[\Delta(z'-z)]dz' \right)\!.
\end{eqnarray}
where $\rho=I/I_c$  is the dimensionless incident light intensity  with $I_c=8\pi^2 c K_2n_e\delta n/[\lambda^2(n_e+n_o)]$ and 
\begin{eqnarray}
\label{phase_del}
\pi\Delta=2\pi L\delta n/\lambda
\end{eqnarray}
is the phase delay between $e$- and $o$-waves through the whole layer with birefringence $\delta n=n_e-n_o$, where $n_e$ and $n_o$ are refractive indices of the $o$ and $e$ waves, respectively. Here we introduce the normalized length $z\rightarrow z \pi/L$ and time $t\rightarrow t/\tau$ where $\tau=\gamma_1 L^2/(\pi^2 K_2)$ is a characteristic relaxation time with the rotational viscosity $\gamma_1$ and the twist Frank elastic constant $K_2$ \cite{Gen}. 

We perform then the linear stability analysis writing $\Phi(z,t)=\Phi(z)\exp(\sigma t)$ and inserting the expansion 
\begin{eqnarray}
\label{set_sin}
\Phi(z,t)=\sum_m\varphi_m(t)\sin(mz).
\end{eqnarray}
into Eq.~(\ref{dir_eq_F_zero}) with further projection on $\varphi_n$ (Galerkin method). The resulting eigenvalue problem is solved then numerically (see \cite{krim_planar} for details). 

The analysis shows that the primary instability of the planar state is a stationary bifurcation only below a critical value       
$\Delta_c=0.64$ of the phase delay and a Hopf bifurcation otherwise, as indicated by the blue dashed line in Fig.~\ref{fig_scatt}. We notice that the symmetry $S$ is spontaneously broken at the threshold in former case. The change from stationary to Hopf bifurcation is accompanied by a jump of the threshold intensity \cite{krim_planar}.

%
\section{Primary instability with realistic light field boundary conditions}
\label{lin_an_sc}
%

As a next step, we consider realistic boundary conditions at $z=0,\pi$, namely an isotropic dielectric material that sandwiches the nematic layer with refractive index $n_{\rm out}$. For convenience, we introduce the labels $i$ and $t$ for the first ($z<0$, incident side) and the second ($z>\pi$, transmitted side) semi-infinite surrounding dielectric material, and $N$ for the nematic layer. The modified linear equation requires a derivation of the expressions for the $o$ and $e$ waves inside nematic to the first order in $\Phi$ that account for backward waves contributions. This is done by using the Berremen's formalism \cite{Berr}, when the Maxwell equations 
are cast in the matrix form for the four component vector $\bar{\Psi}^T=(E_x\,, H_y\,,E_y\,,-H_x)$.  Here ${\bf E}(z,t)$, ${\bf H}(z,t)$ are the amplitudes of the electric and magnetic fields, ${\bf E}({\bf r},t)=1/2({\bf E}(z,t) e^{-i\omega t}+c.c.)$,  ${\bf H}({\bf r},t)=1/2({\bf H}(z,t) e^{-i\omega t}+c.c.)$, that vary slowly in time compared to $\omega^{-1}$. The vector $\bar{\Psi}$ then can be written as a superposition of the four proper electromagnetic waves (two forward- and two backward-propagating) propagating across the layer ($k_x=k_y=0$) with the same value $\omega$ for $i$, $t$ and $N$ media (Oldano's formalism \cite{Oldano}), namely
\begin{eqnarray}
\label{eq_expn}
\bar{\Psi}=\sum_{j=1}^4 f_j \psi^{j}=
f_1 \psi^{1}+f_2 \psi^{2}+f_3 \psi^{3}+f_4 \psi^{4}=
{\bf\sf T}\phi,
\end{eqnarray}
where ${\bf\sf T}$ is the matrix whose columns are eigenvectors $\psi^j$ and $\phi$ is the vector of amplitudes
$\phi^T=(f_1,\,f_2,\,f_3,\,f_4)$. 

The vector of amplitudes $\phi$  at the left side of the first boundary plane ($z=0^-$) and at the right side of the second boundary plane ($z=\pi^+$)  can be written as
\begin{eqnarray}
\label{phi0min}
\phi^T(0^-)=(0,\,a_0,\,r_e,\,r_o),\, \phi^T(\pi^+)=(t_e,\,t_o,\,0,\,0)
\end{eqnarray}
The scattering problem consists of expressing the amplitudes $r_e$, $r_o$, $t_e$, $t_o$ of the reflected and transmitted $e$- and $o$ waves as a function of the amplitude $a_0$ of the incident $o$-wave. The continuity of the tangential components at $z=0,\pi$ of the vectors ${\bf E}$ and ${\bf H}$ requires 
\begin{eqnarray}
\bar{\Psi}(0^-)=\bar{\Psi}(0^+),\,\,\,\bar{\Psi}(\pi^-)=\bar{\Psi}(\pi^+).
\end{eqnarray}
so that we can write
\begin{eqnarray}
\label{bc_1}
\!\!\!{\bf\sf T}_i\phi(0^-)={\bf\sf T}_N(0^+)\phi(0^+),{\bf\sf T}_N(\pi^-)\phi(\pi^-)={\bf\sf T}_t\phi(\pi^+),
\end{eqnarray}
where ${\bf\sf T}_i$ and ${\bf\sf T}_t$ are as follows
\begin{eqnarray}
\label{Ti_matrix}
{\bf\sf T}_i={\bf\sf T}_t=\dfrac{1}{\sqrt{2}}
\begin{pmatrix}
\dfrac{1}{\sqrt{n_i}}&&0&&\dfrac{1}{\sqrt{n_i}}&&0
\\
\sqrt{n_i}&&0&&-\sqrt{n_i}&&0
\\
0&&\dfrac{1}{\sqrt{n_i}}&&0&&\dfrac{1}{\sqrt{n_i}}
\\
0&&\sqrt{n_i}&&0&&-\sqrt{n_i}
\end{pmatrix}.
\end{eqnarray}
Due to the boundary conditions for the director we have ${\bf\sf T_N}(0^+)={\bf\sf T_N}(\pi^-)\equiv {\bf\sf T_N^{0}}$ given by
\begin{eqnarray}
\label{TN_matrix}
{\bf\sf T}_N^{0}=\dfrac{1}{\sqrt{2}}
\begin{pmatrix}
\dfrac{1}{\sqrt{n_e}}&&0&&\dfrac{1}{\sqrt{n_e}}&&0
\\
\sqrt{n_e}&&0&&-\sqrt{n_e}&&0
\\
0&&\dfrac{1}{\sqrt{n_o}}&&0&&\dfrac{1}{\sqrt{n_o}}
\\
0&&\sqrt{n_o}&&0&&-\sqrt{n_o}
\end{pmatrix}.
\end{eqnarray}
The evolution of the electromagnetic field inside nematic can be written by introducing the propagation matrix ${\bf\sf K}(\pi)$ as
\begin{eqnarray}
\label{evol_1}
\phi(\pi^-)={\bf\sf K(\pi)}\phi(0^+),
\end{eqnarray}
where
\begin{eqnarray}
\label{K_matrix_calc}
\!\!\!\!\!\!\!\!\!
{\bf\sf  K}(\pi)=i \alpha \bar{k}_0e^{i \pi \bar{k}_0 n_o}\times
\begin{pmatrix}
\dfrac{e^{i \pi \bar{k}_0 d_1}}{i \alpha \bar{k}_0}&\xi_{1} e^{i \pi \Delta}&0
&\xi_{2} e^{i \pi \Delta }
\\
\xi_{1}&\dfrac{1}{i \alpha \bar{k}_0}&-\xi_{2} &0
\\
0&-\xi_{2}^\star e^{-i \pi q\Delta}&\dfrac{e^{-i \pi q\Delta}}{i \alpha \bar{k}_0}&
-\xi_{1}^\star e^{-i \pi q\Delta}
\\
-\xi_{2}^\star e^{i \pi \Delta (1-q)}&0&
-\xi_{1} e^{i \pi \Delta (1-q)}&\dfrac{e^{i \pi \Delta (1-q)}}{i \alpha \bar{k}_0}
\end{pmatrix}.
\end{eqnarray}
Here
\begin{eqnarray}
\label{q_def}
\alpha=\dfrac{n_e^2-n_o^2}{2\sqrt{n_en_o}},\,\,q=\dfrac{n_e+n_o}{\delta n},
\end{eqnarray}
and
\begin{eqnarray}
\xi_{1}=\int_0^{\pi}e^{-i \Delta z}\Phi(z) dz,\,\,\,\xi_{2}=\int_0^{\pi}e^{-i \Delta q z}\Phi(z) dz.
\end{eqnarray}
Finally, a set of four independent equations that allows to express $r_e$, $r_o$, $t_e$ and $t_o$ as a function of $a_0$ is derived
\begin{eqnarray}
\label{eq_tr_ref}
\phi(0^-)={\bf\sf T_i^{-1}}{\bf\sf T}_N^{0}{\bf\sf K^{-1}(\pi)}({\bf\sf T}_N^{0})^{-1}{\bf\sf T}_t\phi(\pi^+).
\end{eqnarray}
In general, the resulting  expressions for the reflected and transmitted waves $r_e,r_o,t_e$ and $t_o$ are  cumbersome, however, it is instructive to consider the particular case $n_{\rm out}=n_o$ that simplifies the algebra a lot and gives
\begin{eqnarray}
\nonumber
r_e&=&\dfrac{ia_0q\Delta \int_0^\pi\left[e^{i\Delta(\pi-z)}+q e^{-i\Delta q (\pi-z)}\right]\Phi(z)\,dz }{q^2e^{-i \pi q\Delta}-e^{i \pi \Delta}},
\\
\label{t2}
r_o&=&0,
\\\nonumber
t_e&=&\dfrac{ia_0q\Delta e^{\frac{i}{2} \pi \Delta(q-1)}\int_0^\pi\left[q e^{-i\Delta z}+e^{i\Delta q z}\right]\Phi(z)\,dz}
{q^2e^{-i \pi q\Delta}-e^{i \pi \Delta}},
\\\nonumber
t_o&=&a_0 e^{-\frac{i}{2} \pi \Delta(q-1)},
\end{eqnarray}
where $q=(n_e+n_o)/\delta n$. As is expected, $r_o=0$ because an incident ordinary wave is perfectly matched ($n_{\rm out}=n_o$). Thus the $o$ wave is entirely transmitted acquiring only the phase according to Eq.~(\ref{t2}).

Next, we determine the light field distribution inside nematic by integrating the equations for the amplitudes $f_i(z)$, since the initial condition $\phi(0^+)=({\bf\sf T_N^0})^{-1}\,{\bf\sf T_i}\,(0,\,a_0,\,r_e,\,r_o)^T$ is found. Finally, the linearized expression  for the electromagnetic term can then be calculated, leading to the following integro-differential equation for $\Phi$ 
\begin{eqnarray}
\label{dir_eq}
\partial_t\Phi=\partial_z^2\Phi+2\rho\Delta^2
\left(\Phi+\Delta\int_0^z\Phi(z')\sin[\Delta(z'-z)]dz'+
\Delta\cdot Im[F(z)] \right),
\end{eqnarray}
where $F(z)$ describes the contribution of the backward waves. Then, the stability analysis procedure is similar to the case of ideal boundary conditions (see Sec.~\ref{subsec_id_bc}). In the particular case $n_{\rm out}=n_o$, the latter function writes explicitly
\begin{eqnarray}
\label{fun_F}
F(z)=\int_0^ze^{-i q\Delta (z-z')}\Phi(z')dz'+
\dfrac{(e^{i\Delta z}+qe^{-i q \Delta z})}{e^{i \pi \Delta}-q^2e^{-i \pi q\Delta}}\cdot
\int_0^\pi\left[e^{i\Delta(\pi-z)}+q e^{-i\Delta q (\pi-z)}\right]\Phi(z)\,dz
\end{eqnarray}

\begin{figure}
\center{\includegraphics[width=0.7\columnwidth ]{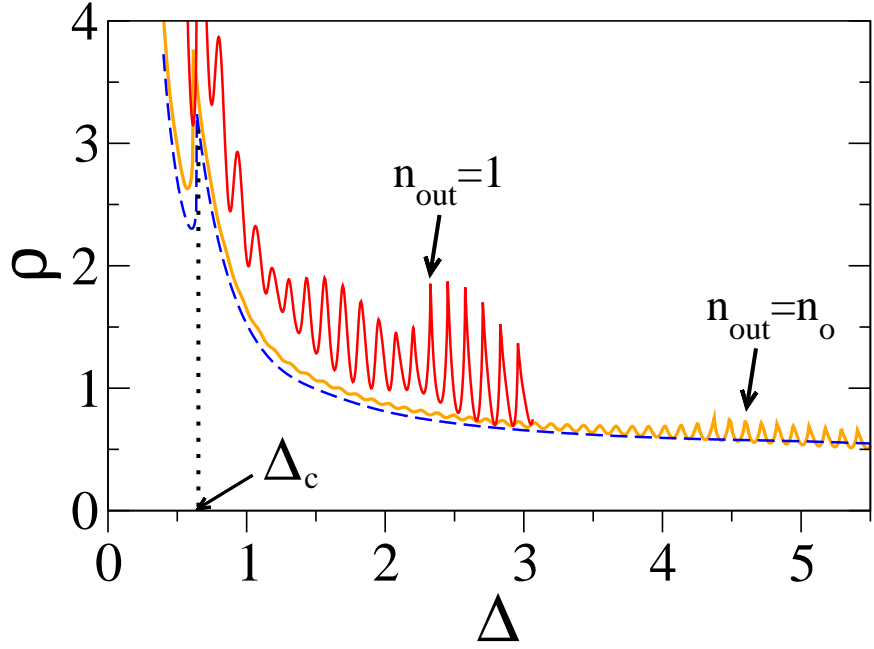}}
\caption{(Color online) Primary instability threshold of the planar state on the $(\Delta,\rho)$ plane with (solid lines) and without (blue dashed line) the effect of backward waves. Orange (light gray) line: refractive index of outer media $n_{\rm out}=n_o$. Red (dark gray) line: $n_{\rm out}=1$. Lines with $\Delta <\Delta_c$ ($\Delta \ge \Delta_c$) correspond to stationary (Hopf) bifurcation.} \label{fig_scatt}
\end{figure}

The results are summarized in Fig.~\ref{fig_scatt} where primary instability threshold is plotted in the $(\Delta,\rho)$ plane with (solid curves, for the two cases $n_{\rm out}=n_o$ and $n_{\rm out}=1$) and without (dashed curve) taking into account the contribution of backward waves generated at the nematic boundaries. The presence of backward waves has a clear qualitative effect, namely, the oscillation of the threshold with respect to the phase delay $\Delta$ of the nematic layer. Such an oscillation is driven by the dependence of the function $F(z)$ on $\Delta$. In the case $n_{\rm out}=n_o$, the resulting oscillation has a period $2/q$ [see Eq.~(\ref{fun_F})]. It corresponds to a period of $\lambda/(n_e+n_o)$  when considering the dependence of the threshold versus thickness $L$. The fact of oscillations itself emphasizes the interference between forward and backward propagating waves to be at the origin of the modulation of the threshold. Accordingly, the modulation depth drastically increases when the refractive index contrast at the nematic boundaries is increased, as shown when taking $n_{out}=1$ (see Fig.~\ref{fig_scatt}). To conclude this section, let us note that in a practical situation the representative case will correspond to $n_{\rm out}=1$. Indeed, a sample is typically made of a nematic slab sandwiched into two glass substrates (with refractive index $n_{\rm glass} \simeq n_o$) surrounded with air, hence we expect the air/glass interfaces to control the modulation of the OFT threshold, which therefore corresponds qualitatively to the case $n_{\rm out}=1$.

\section{Conclusions}

We report on the effect of back- and forward scattering on the Fr\'eedericksz transition in a nematic layer with planar alignment. The light intensity and the phase delay induced by the layer are the control parameters of the system. 
We demonstrate that the primary threshold exhibits oscillations with the phase delay. This effect is especially important when the jump of refractivity index at the boundaries of nematic layer occurs.


%

%

\newpage


\end{document}